\documentclass[english,pra,preprint,showpacs]{revtex4-1}
\usepackage[T1]{fontenc}
\usepackage[latin9]{inputenc}
\usepackage{amsmath}
\usepackage{amssymb}
\usepackage{graphicx}
\usepackage{esint}
\usepackage{babel}
\begin{document}

\title{Engineering frequency-dependent superfluidity in Bose-Fermi mixtures}

\author{Maksims Arzamasovs}

\affiliation{Wilczek Quantum Center, Zhejiang University of Technology, Hangzhou,
Zhejiang, China}

\affiliation{University of Pittsburgh, Pittsburgh, PA, U.S.A.}

\author{Bo Liu}

\affiliation{Department of Applied Physics, School of Science, Xi'an Jiaotong
University, Xi'an 710049, Shaanxi, China}

\affiliation{Shaanxi Province Key Laboratory of Quantum Information and Quantum
Optoelectronic Devices, Xi\textquoteright an Jiaotong University,
Xi\textquoteright an 710049, Shaanxi, China}

\date{31 July 2017}
\begin{abstract}
Unconventional superconductivity or superfluidity are among the most
exciting and fascinating quantum states in condensed matter physics.
Usually these states are characterized by non-trivial spatial symmetry
of the pairing order parameter, such as in $^{3}He$ and high-$T_{c}$
cuprates. Besides spatial dependence the order parameter could have
unconventional frequency dependence, which is also allowed by Fermi-Dirac
statistics. For instance, odd-frequency pairing is an exciting paradigm
when discussing exotic superfluidity or superconductivity and is yet
to be realized in the experiments. In this paper we propose a symmetry-based
method of controlling frequency dependence of the pairing order parameter
via manipulating the inversion symmetry of the system. First, a toy
model is introduced to illustrate that frequency dependence of the
order parameter can be adjusted by controlling the inversion symmetry
of the system. Second, taking advantage of the recent rapid developments
of shaken optical lattices in ultracold gases, we propose a Bose-Fermi
mixture to realize such frequency dependent superfluids. The key idea
is introducing the frequency-dependent attraction between Fermions
mediated by Bogoliubov phonons with asymmetric dispersion. Our proposal
should pave an alternative way for exploring frequency-dependent superconductors
or superfluids with cold atoms.
\end{abstract}


\maketitle

\section{Introduction}

Symmetry plays an important role in physics of superconductors or
superfluids and influences their properties in a profound way. For
example, research on unconventional superconductors where the pairing
order parameter has non-trivial spatial symmetry, such as spin-singlet
$d$-wave in cuprates and spin-triplet $p$-wave in $Sr_{2}RuO_{4}$
has generated tremendous interest in strongly correlated electron
systems. In addition to spatial dependence the order parameter may
also depend on frequency. Indeed, in models describing the electron-electron
attraction induced by interaction with lattice phonons (Frohlih Hamiltonian)
the effective electron-electron attraction, and therefore the gap,
are frequency-dependent \citep{Mahan}. The gap happens to be even
in frequency but this does not have to be the case in general. In
fact the idea of pairing order parameter with odd frequency dependence
was proposed by V.L. Berezinskii in 1974 \citep{Berezinskii}. This
Berezinskii conjecture has attracted considerable attention, in particular
with respect to some exotic phenomena like the anomalous proximity
effect in superconductor-ferromagnet junctions \citep{Nagaosa}.

However the question of controlling the frequency dependence of the
pairing order parameter remains unresolved. The idea of symmetry breaking
may offer a clue: indeed, it has been skillfully used recently to
obtain exotic superconducting and superfluid ordering in the ultracold
atomic setting \citep{IsaacssonGirvin,LiuWu,HemmerichSmith}. In this
paper we report on a method of engineering frequency-dependent superconductors
or superfluids based on manipulating the inversion symmetry. The main
idea can be understood through the following symmetry argument. The
superconducting order parameter is defined by a non-vanishing value
of the anomalous correlator

\begin{equation}
F(\alpha,\alpha')=\left\langle \psi(\alpha)\psi(\alpha')\right\rangle ,\label{eq:DefinitionOrderParameter}
\end{equation}
where $\alpha,\alpha'$ stand for the internal degrees of freedom,
such as coordinate, time, spin and band indices of the Fermi field
$\psi$ and $\left\langle ...\right\rangle $ stands for the ground
state or thermal state average. Let us assume that fermions pair up
in a spin-singlet state (as is the case for the BCS model \citep{BCS}).
According to Fermi-Dirac statistics $F(\alpha,\alpha')$ should be
antisymmetric with respect to exchange of $\alpha$ and $\alpha'$.
Spin-singlet state is already antisymmetric therefore the remaining
part of the anomalous correlator should satisfy
\begin{equation}
F(q,\omega_{n})=F(-q,-\omega_{n}),\label{eq:SingletSymmetry}
\end{equation}
where $q$ and $\omega_{n}$ are the momentum and Matsubara frequency.
In order to control the frequency dependence of the pairing we suggest
manipulating the inversion symmetry of the system. Indeed, if the
system is inversion-symmetric then
\begin{equation}
F(q,\omega_{n})=F(-q,\omega_{n}),\label{eq:EvenFreq1}
\end{equation}
which according to Eq. (\ref{eq:SingletSymmetry}) makes it also even
in frequency, 
\begin{equation}
F(q,\omega_{n})=F(q,-\omega_{n}).\label{eq:EvenFreq2}
\end{equation}
By introducing the inversion symmetry breaking the odd-frequency superconductivity
\[
F(q,\omega_{n})=-F(-q,\omega_{n}),
\]
\begin{equation}
F(q,\omega_{n})=-F(q,-\omega_{n})\label{eq:OddFreq}
\end{equation}
may be feasible. In the following we will first use a toy model to
show how this idea works. Then a cold atom based experiment is proposed.

\section{Toy model}

Let us first consider a 1D toy model of a Bose-Fermi mixture:
\begin{equation}
\hat{H}=\hat{H}_{F}+\hat{H}_{B}+\hat{H}_{BF}.\label{eq:ToyModelHamiltonian}
\end{equation}
Here for simplicity we assume that $\hat{H}_{F}$ describes non-interacting
spin-1/2 fermions,
\begin{equation}
\hat{H}_{F}=\sum_{k,\sigma}\left(\epsilon_{F}(k)-\mu_{F}\right)\hat{f}_{k,\sigma}^{\dagger}\hat{f}_{k,\sigma},\label{eq:FreeSpinlessFerm}
\end{equation}
and $\hat{H}_{B}$ describes spinless bosons,
\begin{equation}
\hat{H}_{B}=\sum_{q}\left(\epsilon_{B}(q)-\mu_{B}\right)\hat{b}_{q}^{\dagger}\hat{b}_{q}.\label{eq:FreeBosons}
\end{equation}
The key idea for controlling the frequency dependence of superconducting
order parameter relies on introducing the inversion symmetry breaking
via asymmetric energy dispersion of bosons, $\epsilon_{B}(-q)\neq\epsilon_{B}(q)$,
that will be illustrated in detail below. Finally, $\hat{H}_{BF}$
describes the fermion-boson interaction,
\begin{equation}
\hat{H}_{BF}=\frac{U_{BF}}{\sqrt{N}}\sum_{q}\hat{\rho}_{q}\left(\hat{b}_{-q}+\hat{b}_{q}^{\dagger}\right)\label{eq:ToyModelInteractions}
\end{equation}
where $\hat{\rho}_{q}=N^{-1}\sum_{k,\sigma}\hat{f}_{k,\sigma}^{\dagger}\hat{f}_{q+k,\sigma}$
stands for the Fourier transform of fermion density and $N$ is the
total number of lattice sites. In this toy model we can integrate
out the bosons exactly using the path integral formalism, see Appendix
A for details, and the resulting induced density-density type interaction
between fermions is given by
\begin{equation}
V_{Ind}(q,\omega_{n})=-\frac{U_{BF}^{2}}{2}\frac{\epsilon_{B}(q)+\epsilon_{B}(-q)-2\mu_B}{\omega_{n}^{2}+\left(\epsilon_{B}(q)-\mu_B\right)\left(\epsilon_{B}(-q)-\mu_B\right)+i\omega_{n}\left(\epsilon_{B}(q)-\epsilon_{B}(-q)\right)},\label{eq:VInd}
\end{equation}
where $\omega_{n}=(2n+1)\pi T$ are Matsubara frequencies and $\hbar=1$,
$k_{B}=1$ throughout the paper.

To demonstrate how the inversion symmetry breaking can be used to
control frequency dependence of the order parameter we assume the
following dispersion for bosons
\begin{equation}
\epsilon_{B}(q)=\begin{cases}
c_{R}\left|q\right|, & q\geq0\\
c_{L}\left|q\right|, & q<0
\end{cases}\label{eq:ToyLinearDispersion}
\end{equation}
as a concrete example. We employ Eliashberg equations \citep{Alexandrov}
to find the superconducting gap $\Delta(k,\omega_{n})$ as a function
of momentum and frequency:
\[
\Delta(k,\omega_{n})=-\frac{T}{N}\sum_{k',n'}\frac{V_{Ind}(k-k',\omega_{n}-\omega_{n'})\Delta(k',\omega_{n'})}{\omega_{n'}^{2}Z^{2}(k',\omega_{n'})+\xi^{2}(k',\omega_{n'})+\left|\Delta(k',\omega_{n'})\right|^{2}},
\]
\begin{equation}
(1-Z(k,\omega_{n}))i\omega_{n}=\frac{T}{N}\sum_{k',n'}\frac{V_{Ind}(k-k',\omega_{n}-\omega_{n'})i\omega_{n'}Z(k',\omega_{n'})}{\omega_{n'}^{2}Z^{2}(k',\omega_{n'})+\xi^{2}(k',\omega_{n'})+\left|\Delta(k',\omega_{n'})\right|^{2}},\label{eq:ElaishbergEquations}
\end{equation}
\[
\chi(k,\omega_{n})=\frac{T}{N}\sum_{k',n'}\frac{V_{Ind}(k-k',\omega_{n}-\omega_{n'})\xi(k',\omega_{n'})}{\omega_{n'}^{2}Z^{2}(k',\omega_{n'})+\xi^{2}(k',\omega_{n'})+\left|\Delta(k',\omega_{n'})\right|^{2}},
\]
where $\xi(k,\omega_{n})=\epsilon_{F}(k)+\chi(k,\omega_{n})$, $Z(k,\omega_{n})$
is the fermion mass renormalization and $\chi(k,\omega_{n})$ the
renormalization of the chemical potential. To simplify the problem,
here we consider 1D case and assume the fermions to have $s$-band-like
dispersion, $\epsilon_{F}(k)=-t_{F}\cos(ka)$, $k$ being
lattice momentum and $a$ the lattice constant. 

As shown in Fig. 1, when bosons have asymmetric dispersion (i.e. $c_{R}/c_{L}\neq1$)
and the inversion symmetry is broken, the odd-frequency component
of the superconducting gap emerges. Furthermore, as shown on Fig.
2, the frequency-dependence of the gap can be controlled by manipulating
the degree of asymmetry, i.e. the ratio $c_{R}/c_{L}$. It shows that
the stronger the boson dispersion asymmetry the more favorable is
the odd-frequency component of the superconducting gap. We note that
the superconducting pairing considered here is in the spin-singlet
channel, since the spin-singlet pairing always has lower free energy
compared to the spin-triplet counterpart, as confirmed numerically. 

\begin{figure}
\includegraphics{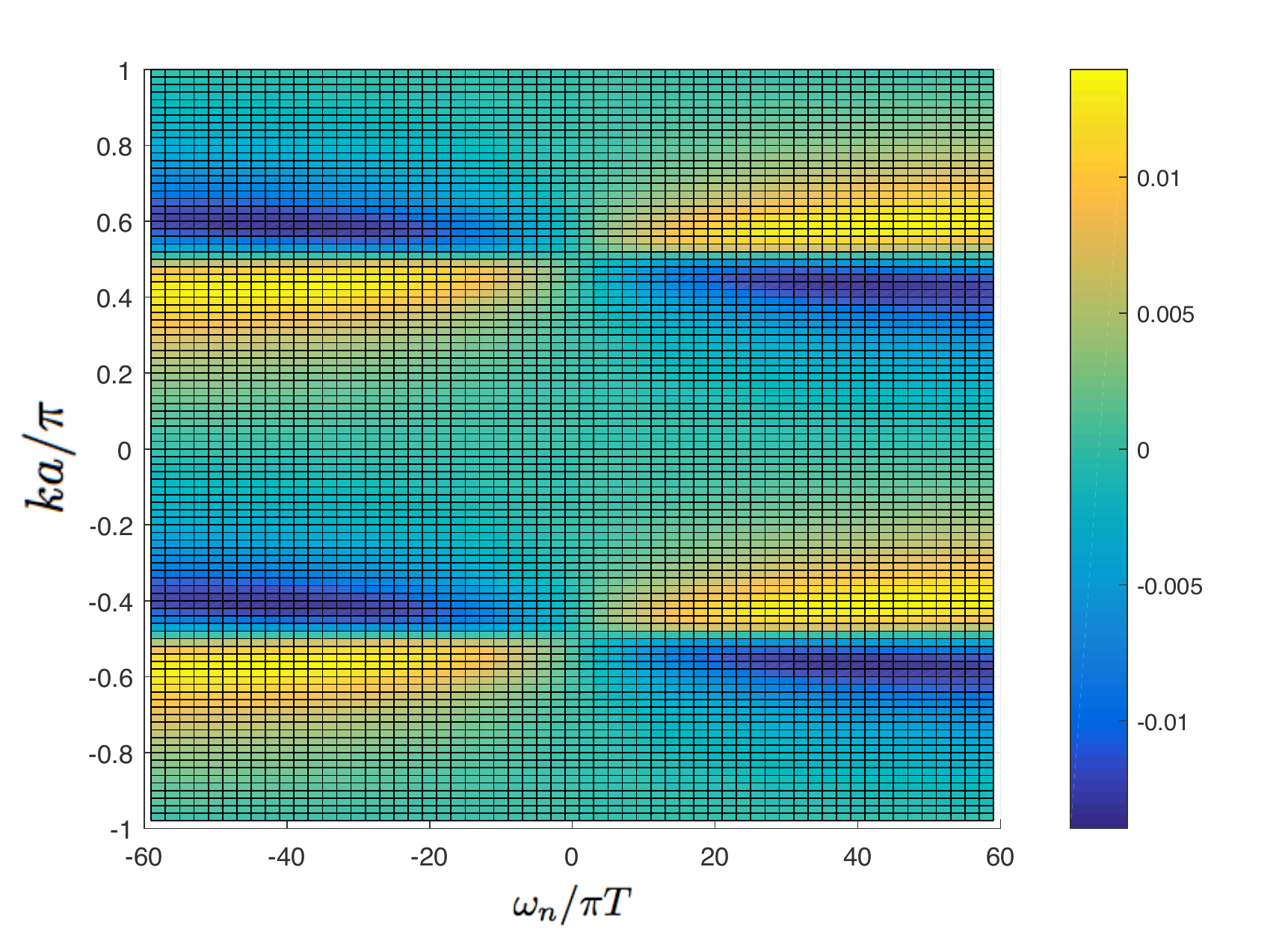}\caption{Odd-frequency component of the superconducting gap, in the units of
$t_{F}$, for the case of non-isotropic dispersion with $c_{R}/c_{L}=3$
and $c_{L}/a=t_{F}$, $\mu_{B}=-0.005t_{F}$, $t_{F}=1$, $\mu_{F}=0$,
$U=t_{F}$, $T=0.001t_{F}$. Lattice momentum is plotted in units
of $\pi/a$, $a$ being the lattice constant. Throughout the paper
$k_{B}=1$, $\hbar=1$.}
\end{figure}

\begin{figure}
\includegraphics[scale=0.7]{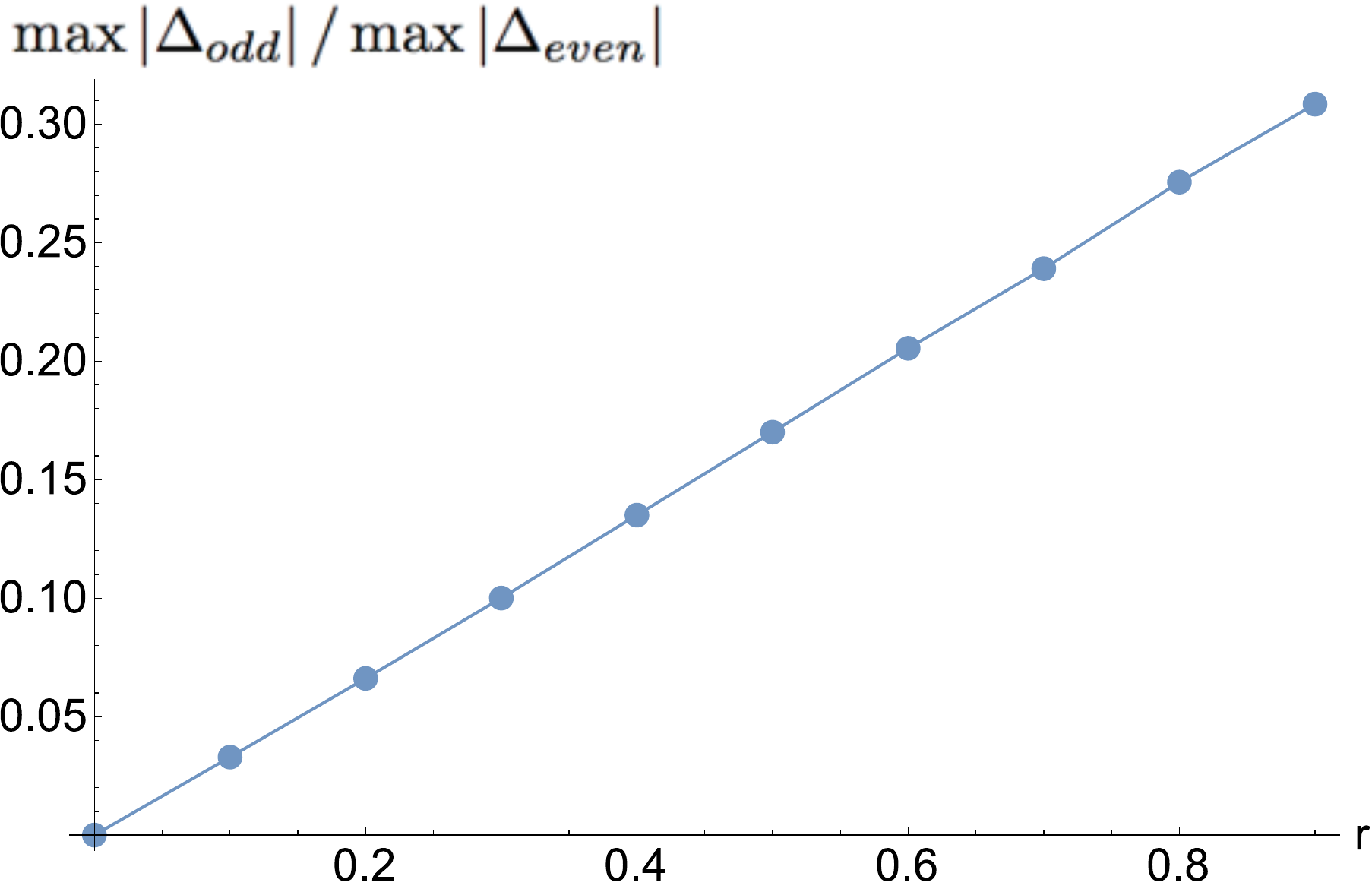}\caption{$\max\left|\Delta_{odd}\right|/\max\left|\Delta_{even}\right|$ plotted
as a function of $r$, where $c_{R}/c_{L}=\frac{1+r}{1-r}$.}

\end{figure}

\section{Superconducting ordering via pairing with shaken bosons}

In the following we explore the possibility of realizing frequency-dependent
superfluidity via a cold-atom based system. Let us consider a Bose-Fermi
mixture in a 1D optical superlattice as shown in Fig. 3 and further
consider the bosons to be in a shaken lattice. Lattice shaking will
change the energy dispersion of bosons to a double-dip shape, see
Appendix B. By further assuming that the bosons are in the plane-wave
state (i.e. that condensation in one of the minima of the double-dip
dispersion has occurred), the induced attraction between fermions
can be obtained as follows.

\begin{figure}
\includegraphics[scale=0.5]{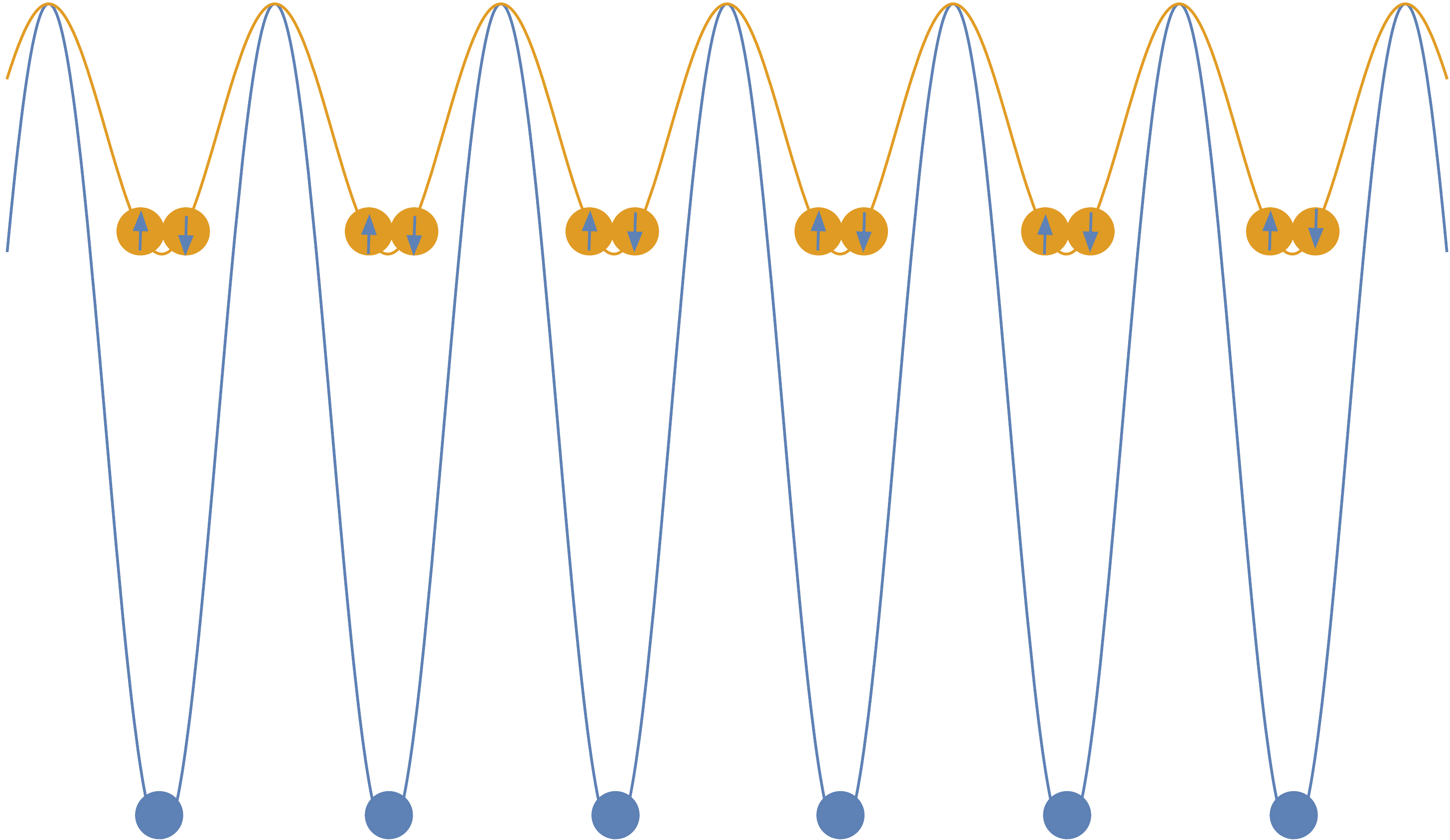}\caption{Boson (blue, deeper lattice) and fermion (orange, shallower lattice)
mixture in a 1D superlattice. The scheme could use alkali-earth atoms
in different orbital states, $^{1}S_{0}$ and $^{3}P_{0}$. The Bose-Fermi
interactions can be tuned using orbital Feshbach resonance techniques,
e.g. see \citep{OrbitalFeshbach1}, \citep{OrbitalFeshbachExp1} and
\citep{OrbitaFeshbachExp2}.}
\end{figure}

We start with the Hamiltonian which is a generalization of the Bose-Hubbard
model where the momentum-space dispersion of free particles $\epsilon_{B}(k)$
has a double-dip (see details in Appendix B):
\begin{equation}
\hat{H}_{B}=\sum_{k}\left(\epsilon_{B}(k)-\mu_{B}\right)\hat{b}_{k}^{\dagger}\hat{b}_{k}+\frac{g}{2N}\sum_{k,k',q}\hat{b}_{k+q}^{\dagger}\hat{b}_{k'-q}^{\dagger}\hat{b}_{k'}\hat{b}_{k},\label{eq:GeneralizedBoseHubbard}
\end{equation}
$k$ is lattice momentum in 1D and $\hat{b}_{k}^{\dagger},\,\hat{b}_{k}$-s
are the bosonic creation/annihilation operators. The fermions are
described by
\begin{equation}
\hat{H}_{F}=\sum_{k,\sigma}\left(\epsilon_{F}(k)-\mu_{F}\right)\hat{f}_{k,\sigma}^{\dagger}\hat{f}_{k,\sigma},\label{eq:ProposalFermions}
\end{equation}
where possible existing repulsion between the fermions is treated
at the mean-field level and is absorbed into the chemical potential.
The interaction between bosons and fermions has the form
\begin{equation}
\hat{H}_{BF}=\frac{U_{BF}}{N}\sum_{q}\hat{n}_{-q}\hat{\rho}_{q}\label{eq:BoseFermiInteractionHamiltonian}
\end{equation}
where $\hat{\rho}_{q}$ has the same definition as given below Eq.
(\ref{eq:ToyModelInteractions}) and, similarly, $\hat{n}_{q}$ is
the bosonic density operator.

We assume that bosons condense in the state with momentum $k_{0}$,
corresponding to one of the minima of the double-dip dispersion $\epsilon_{B}(k)$.
The effective attraction between fermions mediated by Bogoliubov phonons
is derived with the help of the path integral formalism. Details are
shown in Appendix C and here we only state the result. The induced
density-density interaction between fermions is given by 
\begin{equation}
V_{Ind}(k,\omega_{n})=-\frac{U_{BF}^{2}n_{0}}{2}\frac{\epsilon_{B}(k_{0}+k)+\epsilon_{B}(k_{0}-k)-2\epsilon_{B}(k_{0})}{(i\omega_{n}+\epsilon_{B}(k_{0}+k)-\epsilon_{B}(k_{0})+gn_{0})(-i\omega_{n}+\epsilon_{B}(k_{0}-k)-\epsilon_{B}(k_{0})+gn_{0})-g^{2}n_{0}^{2}},\label{eq:InducedFermFerm}
\end{equation}
where $n_{0}=N_{0}/N$ is the dimensionless ``density'' of the condensate
($N_{0}$ is the number of bosons in the condensate, $N$ is the total
number of lattice sites, as before). 
\begin{figure}
\includegraphics[scale=0.25]{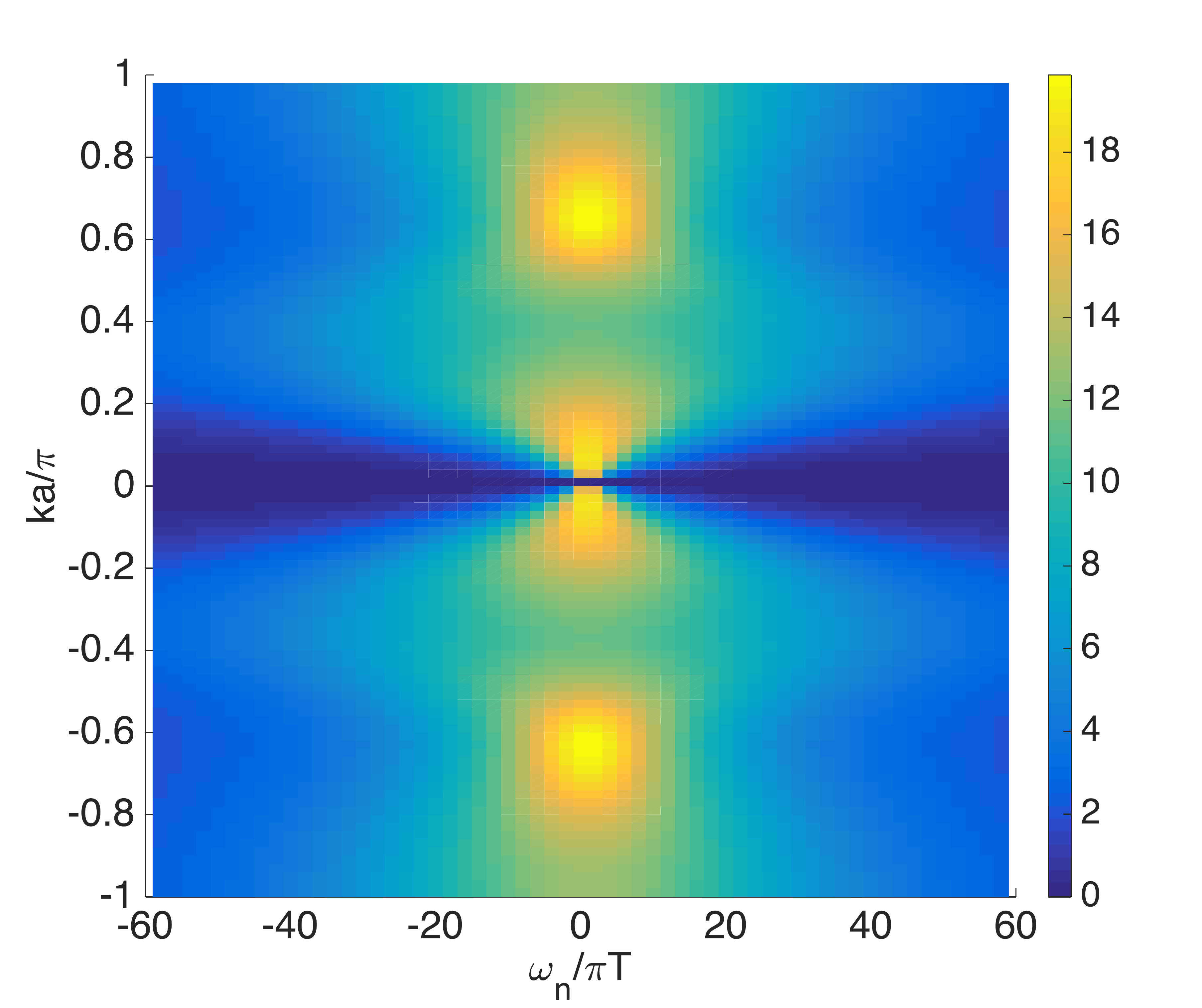}\includegraphics[scale=0.25]{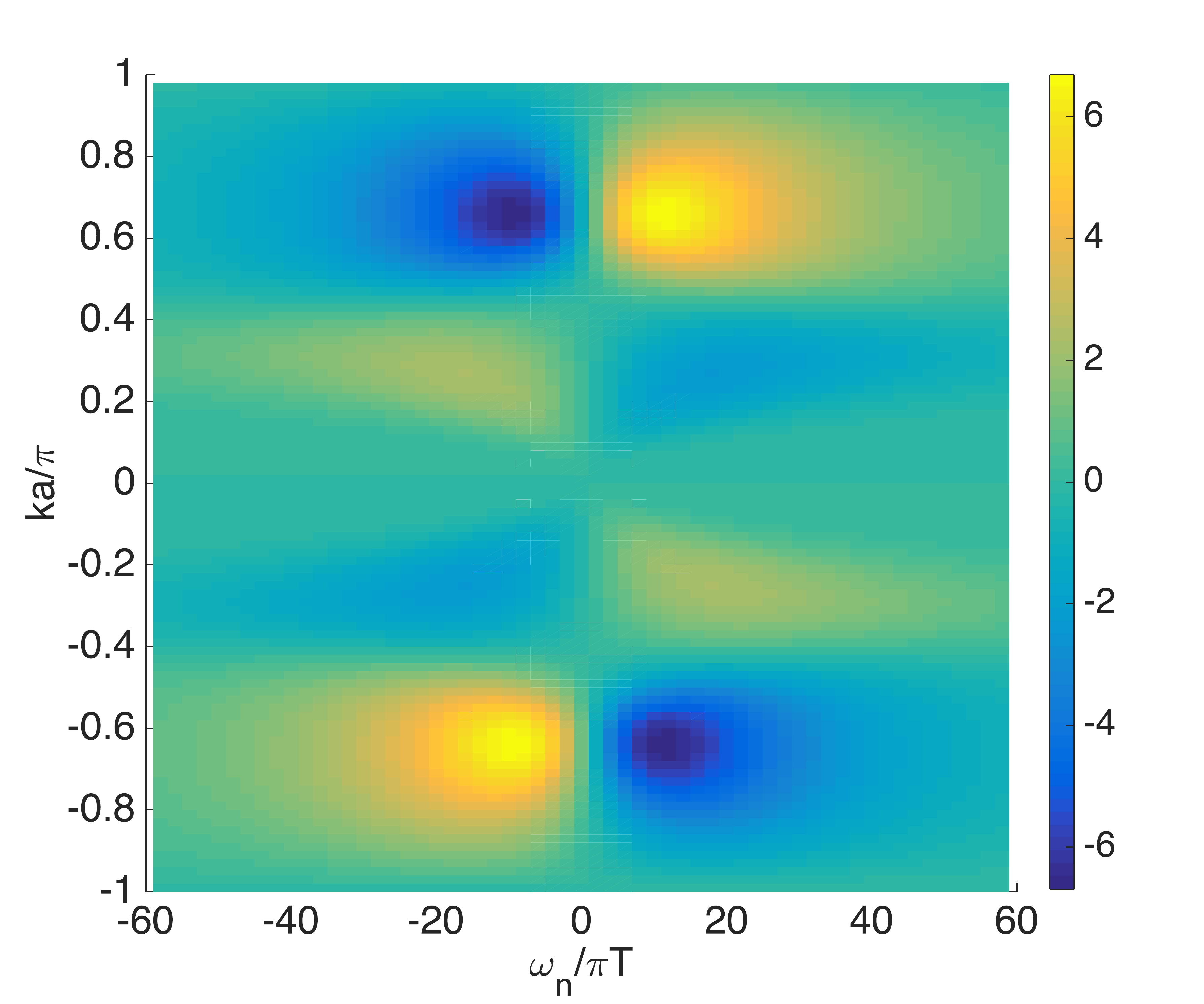}\caption{Strength of induced fermion-fermion attraction potential (in the units
of $E_{R}$), as shown in Eq. (\ref{eq:InducedFermFerm}), $\Re V_{Ind}(k,\omega_{n})$
(right) and $\Im V_{Ind}(k,\omega_{n})$ (left). The shaken bosons
have lattice depth $V=3.5E_{R}$ and shaking frequency $\Omega=5E_{R}$,
where $E_{R}$ is the recoil energy of boson lattice. The boson interaction
strength is such that $gn_{0}=0.05E_{R}$ and $T=0.001E_{R}$.}
\end{figure}

We now proceed to solve the Eliashberg equations for the superconducting
order parameter and accompanying quantities but this time with the
attraction given by Eq. (\ref{eq:InducedFermFerm}). Below we show
sample order parameters calculated based on the inter-fermion attraction
plotted on Fig. 4. As shown on Fig. 5 the frequency-dependent superfluid
is obtained. We find that the real and imaginary parts of the order
parameter have even- and odd-frequency dependence, respectively. This
is in line with the structure of induced attraction Eq. (\ref{eq:InducedFermFerm}).
Namely, $\Re V_{Ind}(k,\omega_{n})$ is even in frequency and $\Im V_{Ind}(k,\omega_{n})$
is odd, see Fig. 4. By tuning the shaking frequency it is possible
to go from the case of symmetric ($\epsilon_{B}(k_{0}+k)=\epsilon_{B}(k_{0}-k)$)
to asymmetric ($\epsilon_{B}(k_{0}+k)\neq\epsilon_{B}(k_{0}-k)$)
Bogoliubov dispersion, and hence to introduce an odd-frequency component
in the superconducting order parameter. 

\begin{figure}
\includegraphics[scale=0.25]{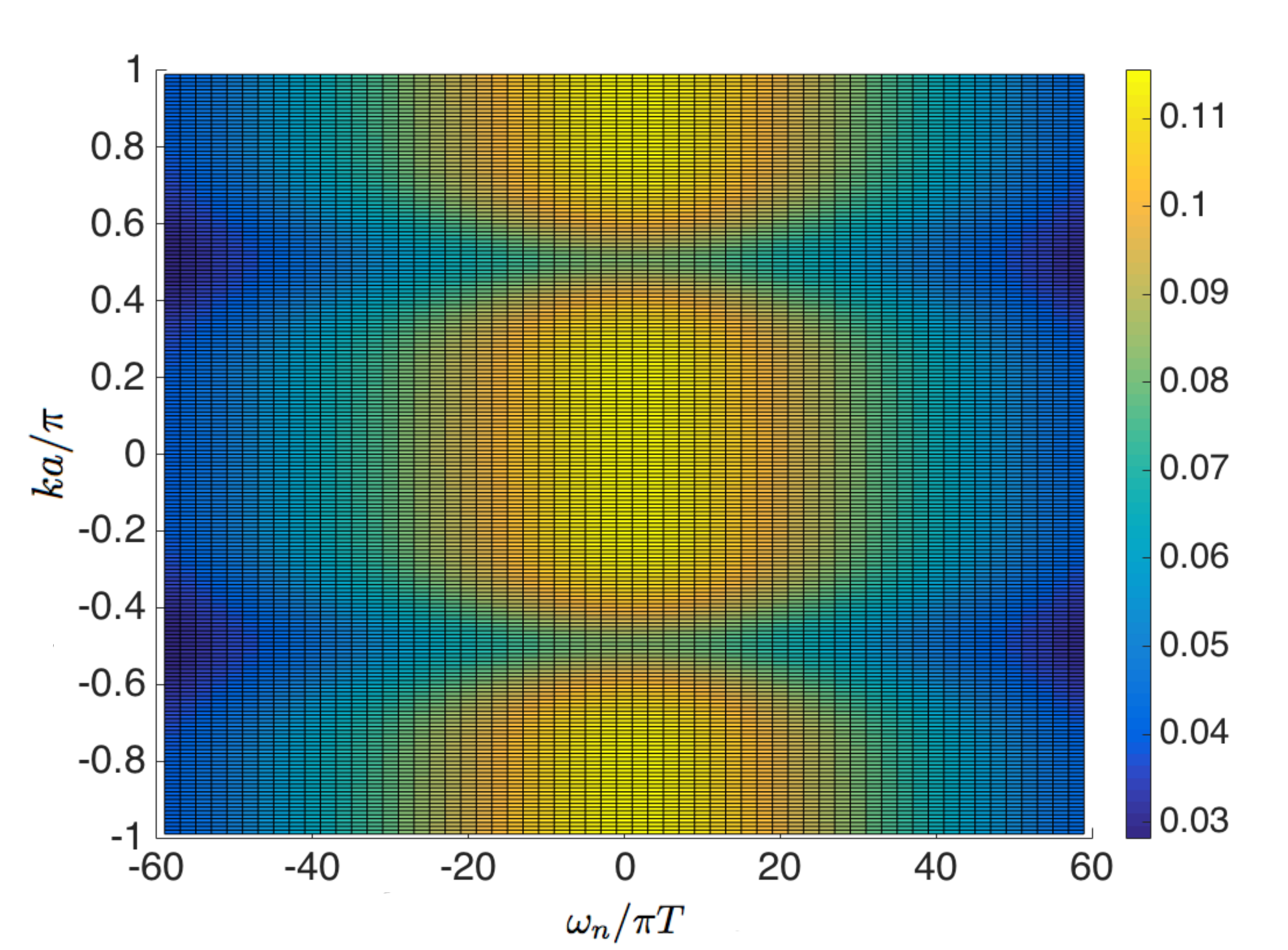}\includegraphics[scale=0.25]{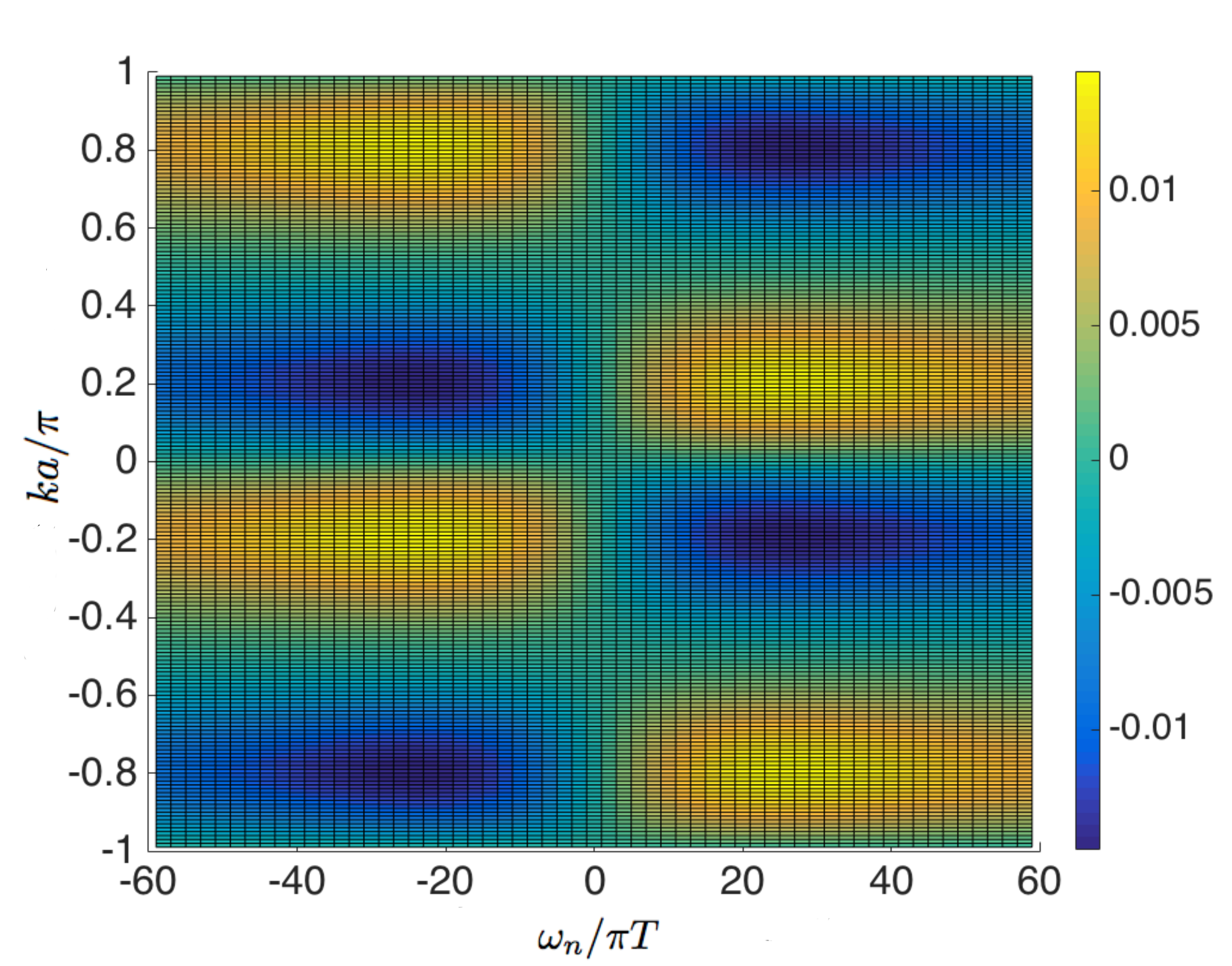}\caption{Real (left) and imaginary (right) parts of the order parameter, in
the units of $t_{F}$, for a non-symmetric interaction potential Eq.
(\ref{eq:InducedFermFerm}). The fermion dispersion $t_{F}=1$, $\mu_{F}=0$,
$U_{BF}=t_{F}$ and $T=0.01t_{F}$. The boson parameters are the same
as in Fig. 4. }
\end{figure}

\section{Conclusions}

We have demonstrated the theoretical feasibility of fermions forming
mixed-frequency superconducting order parameter by means of interacting
with bosons with non-isotropic dispersion. To achieve such non-isotropy
we considered bosons condensing in one of the minima of the double
dip dispersion formed by (quasi) 1D shaking. We have shown that by
tuning the aforementioned parameters it is possible for the superconducting
order parameter to be 10\% odd-frequency or greater. As a final note
we want to mention that shaking the lattice is not the only possibility
for creating non-isotropic and/or the double-dip dispersion, see for
example recent work on spin-orbit coupled BEC \citep{SpielmanReview},
and our scheme can be perfected and/or modified to achieve realization
of a mixed-frequency superconductor in ultracold atomic settings.
\begin{acknowledgments}
The authors would like to thank W. V. Liu and H. Xiong for discussions
and hospitality and X. Lin and X. Yue for discussions.
\end{acknowledgments}

\appendix

\section{Integrating out the toy model bosons}

Toy model Hamiltonian Eq. (\ref{eq:ToyModelHamiltonian}) is quadratic
in bosonic operators. This makes integrating out bosons a straightforward
exercise in Gaussian integration. Using the path integral formalism
we go from Hamiltonian
\begin{equation}
\hat{H}=\sum\left(\epsilon_{B}(q)-\mu_{B}\right)\hat{b}_{q}^{\dagger}\hat{b}_{q}+\frac{U_{BF}}{\sqrt{N}}\sum\hat{\rho}_{q}\left(\hat{b}_{-q}+\hat{b}_{q}^{\dagger}\right)\label{eq:BosonicPart}
\end{equation}
to partition function
\begin{equation}
Z=\int D\bar{b}Db\,\exp\left(-\sum\intop_{0}^{\beta}d\tau\,\bar{b}_{q}(\tau)\left(\partial_{\tau}+\epsilon_{B}(q)-\mu_{B}\right)b_{q}(\tau)-\frac{U_{BF}}{\sqrt{N}}\sum\intop_{0}^{\beta}d\tau\,\rho_{q}(\tau)\left(b_{-q}(\tau)+\bar{b}_{q}(\tau)\right)\right)\label{eq:ToyPartitionFunction}
\end{equation}
where 
\begin{equation}
D\bar{b}Db=\prod d\bar{b}_{q}db_{q}\label{eq:Measure}
\end{equation}
up to an arbitrary normalization factor and $\beta=1/T$ is the inverse
temperature (throughout this paper $k_{B}=1$, $\hbar=1$). Instead
of operators, $\bar{b}_{q}(\tau)$, $b_{q}(\tau)$-s now stand for
complex-valued bosonic fields and $\rho_{q}(\tau)$-s are fields corresponding
to Fourier components of fermion density. In Eq. (\ref{eq:ToyPartitionFunction})
$\rho_{q}(\tau)$-s are treated as external fields. 

We will now integrate out the bosons and obtain the induced interaction
between the fermions. To this end we employ the Matsubara frequency
representation, $b_{q}(\tau)=\sqrt{T}\sum_{n}b_{q}(\omega_{n})e^{-i\omega_{n}\tau}$,
$\bar{b}_{q}(\tau)=\sqrt{T}\sum_{n}\bar{b}_{q}(\omega_{n})e^{i\omega_{n}\tau}$
and $\rho_{q}(\tau)=T\sum_{n}\rho_{q}(\omega_{n})e^{-i\omega_{n}\tau}$,
where $\omega_{n}=2n\pi T$, $n$ being integers. We re-express the
imaginary time integrals in Eq. (\ref{eq:ToyPartitionFunction}) as
sums over Matsubara frequencies $\omega_{n}$,
\begin{equation}
Z=\int D\bar{b}Db\,\exp\left(-\sum_{q,n}\bar{b}_{q}(\omega_{n})\left(-i\omega_{n}+\epsilon_{B}(q)-\mu_{B}\right)b_{q}(\omega_{n})-\frac{U_{BF}\sqrt{T}}{\sqrt{N}}\sum_{q,n}\rho_{q}(\omega_{n})\left(b_{-q}(-\omega_{n})+\bar{b}_{q}(\omega_{n})\right)\right),\label{eq:ToyPartitionFunction-1}
\end{equation}
then integrate out $\bar{b}_{q}(\omega_{n})$, $b_{q}(\omega_{n})$
with the help of the following identity:
\begin{equation}
\int d\bar{z}dz\,e^{-\bar{z}G^{-1}z+\bar{J}z+\bar{z}J}=\frac{e^{\bar{J}GJ}}{\det G^{-1}}.\label{eq:GaussianIntegration}
\end{equation}
This results in the following induced fermion-fermion interaction
action
\begin{equation}
S_{Ind}=-\frac{U_{BF}^{2}T}{2N}\sum_{q,n}\rho_{-q}(-\omega_{n})\frac{\epsilon_{B}(q)+\epsilon_{B}(-q)-2\mu_{B}}{\omega_{n}^{2}+\left(\epsilon_{B}(q)-\mu_{B}\right)\left(\epsilon_{B}(-q)-\mu_{B}\right)+i\omega_{n}\left(\epsilon_{B}(q)-\epsilon_{B}(-q)\right)}\rho_{q}(\omega_{n}),\label{eq:FermFermInteraction}
\end{equation}
from which $V_{Ind}(k,\omega_{n})$ Eq. (\ref{eq:VInd}) is obtained.
The minus sign in front of the sum in Eq. (\ref{eq:FermFermInteraction})
ensures that induced interaction is attractive.

\section{Shaken lattice dispersion}

Here we outline the physics of shaken lattices \citep{ChinShakenLattice}.
Consider two lowest bands ($s-$ and $p-$) of a 1D optical lattice
and their hybridization when the shaking frequency matches the band
gap. The Hamiltonian is
\begin{equation}
H(t)=\frac{k_{x}^{2}}{2m}+V\cos^{2}\left(k_{0}x+\frac{\theta(t)}{2}\right),\label{eq:BasicShakenHamiltonian}
\end{equation}
where $\theta(t)=f\cos(\Omega t)$ describes the periodic shaking
of the lattice. Because of its special form the time-dependent Hamiltonian
Eq. (\ref{eq:BasicShakenHamiltonian}) can be expanded using the properties
of Bessel functions. Write the potential term
\begin{equation}
V\cos^{2}\left(k_{0}x+\frac{f}{2}\cos(\Omega t)\right)=\frac{V}{2}\left(1+\cos\left(2k_{0}x+f\cos(\Omega t)\right)\right)\label{eq:PotentialTerm}
\end{equation}
as
\begin{equation}
\cos\left(2k_{0}x+f\cos(\Omega t)\right)=\frac{1}{2}\left(e^{i2k_{0}x}e^{if\cos(\Omega t)}+e^{-i2k_{0}x}e^{-if\cos(\Omega t)}\right),\label{eq:ExpandedPotentialTerm}
\end{equation}
then use the Jacobi-Anger expansion
\begin{equation}
e^{iz\cos(\phi)}=\sum_{n=-\infty}^{\infty}i^{n}J_{n}(z)e^{in\phi},\label{eq:JacobiAngerExpansion}
\end{equation}
where $J_{n}(z)$ are the $n-$th order Bessel functions of the first
kind. Applying the expansion and using the symmetry of Bessel functions,
\begin{equation}
J_{n}(-z)=(-1)^{n}J_{n}(z),\label{eq:BesselSymmetry}
\end{equation}
\begin{equation}
e^{if\cos(\Omega t)}=\sum_{n=-\infty}^{\infty}i^{n}J_{n}(f)e^{in\Omega t},\label{eq:JacobiAngerAnother}
\end{equation}
\begin{equation}
e^{-if\cos(\Omega t)}=\sum_{n=-\infty}^{\infty}(-i)^{n}J_{n}(f)e^{in\Omega t}\label{eq:BesselExpansion}
\end{equation}
end up with the following expansion for the driving term:
\begin{equation}
\cos\left(2k_{0}x+f\cos(\Omega t)\right)=J_{0}(f)\cos(2k_{0}x)-2J_{1}(f)\sin(2k_{0}x)\sin(\Omega t)+...\label{eq:PotentialTermExpanded}
\end{equation}
where terms with frequencies $2\Omega t$ and higher have been neglected.
Therefore, up to a constant energy shift, the shaken Hamiltonian takes
the form
\begin{equation}
H(t)=\frac{k_{x}^{2}}{2m}+VJ_{0}(f)\cos^{2}(k_{0}x)-VJ_{1}(f)\sin(2k_{0}x)\sin(\Omega t)+...\label{eq:ShakenHamiltonianExpanded}
\end{equation}
The above exercise in algebra allows us to connect physics of shaken
lattice and band mixing. The $s-$ and $p-$ bands of the time-averaged
Hamiltonian
\begin{equation}
H_{av}=\frac{1}{T}\intop_{0}^{T}H(t)dt=\frac{k_{x}^{2}}{2m}+VJ_{0}(f)\cos^{2}(k_{0}x)\label{eq:TimeAveragedHamiltonian}
\end{equation}
mix under the influence of periodic perturbation 
\begin{equation}
V(t)=-VJ_{1}(f)\sin(2k_{0}x)\sin(\Omega t)+...\label{eq:PeriodicPerturbation}
\end{equation}
when the shaking frequency matches the inter-band spacing, forming
two Floquet bands. It was shown in \citep{ChinShakenLattice} that
for a certain parameter window of $V$, $f$, $\Omega$ one of the
Floquet bands possess two degenerate minima at $k_{min},-k_{min}\neq0,\pi$
(e.g. see Fig. 1b in \citep{ChinShakenLattice}). 

To find Floquet bands of the shaken lattice we can follow \citep{ChinShakenLattice}
and project the periodic driving term Eq. (\ref{eq:PeriodicPerturbation})
on the lowest two bands ($s-$ and $p-$) of the time-averaged Hamiltonian
Eq. (\ref{eq:TimeAveragedHamiltonian}) which results in
\begin{equation}
V(t)=-VJ_{1}(f)\left(\begin{array}{cc}
0 & C(k,V,f)\\
C^{*}(k,V,f) & 0
\end{array}\right)\sin(\Omega t)\label{eq:DrivingTerm}
\end{equation}
where $C(k)=\left\langle \Psi_{sk}\right|\sin(2k_{0}x)\left|\Psi_{pk}\right\rangle $
and the diagonal terms vanish due to symmetry considerations. $\left|\Psi_{sk}\right\rangle $
and $\left|\Psi_{pk}\right\rangle $ are the Bloch waves corresponding
to the $s-$ and $p-$ bands of the cosine lattice and can be either
computed numerically using the plane wave expansion or expressed exactly
in terms of Mathieu functions \citep{NIST}. When shaking frequency
$\Omega$ is close to the value of the band gap (the resonance condition)
it is justified to use the rotating wave approximation \citep{Shirley}
which results in the following approximate Floquet Hamiltonian: 
\begin{equation}
H(k)=\left(\begin{array}{cc}
E_{p}(k)-\frac{\Omega}{2} & -\frac{iVJ_{1}(f)C(k)}{2}\\
\frac{iVJ_{1}(f)C^{*}(k)}{2} & E_{s}(k)+\frac{\Omega}{2}
\end{array}\right),\label{eq:RWAHamiltonian}
\end{equation}
where $E_{s}(k)=\left\langle \Psi_{sk}\right|H_{av}\left|\Psi_{sk}\right\rangle $
and similarly for $E_{p}(k)$. All the terms in Eq. (\ref{eq:RWAHamiltonian})
can be obtained numerically exactly in terms of Mathieu functions
and by diagonalizing Eq. (\ref{eq:RWAHamiltonian}) the emergent shaken
(Floquet) bands can be found. Including the inter-boson interactions
will result in Bose condensation around either $k_{min}$ or $-k_{min}$
and the spectrum of excitations above that state (the Bogoliubov phonons)
will end up not being inversion-symmetric \citep{ChinMaxonRoton}.

\section{Integrating out the shaken bosons}

The bosonic part of the partition function takes the form
\begin{equation}
Z=\intop D\bar{b}Db\,\exp\left[-S_{B}-S_{BF}\right],\label{eq:BosonicAction}
\end{equation}
where
\[
S_{B}+S_{BF}=\intop_{0}^{\beta}d\tau\,\sum_{k}\bar{b}_{k}(\tau)\left(\partial_{\tau}+\epsilon_{B}(k)-\mu_{B}\right)b_{k}(\tau)+\frac{g}{2N}\intop_{0}^{\beta}d\tau\,\sum_{k,k',q}\bar{b}_{k+q}\bar{b}_{k'-q}b_{k'}b_{k}+
\]
\begin{equation}
+U_{BF}\intop_{0}^{\beta}d\tau\,\sum_{i}\bar{b}_{i}b_{i}\rho_{i}\label{eq:ActualBosonicAction}
\end{equation}
and 
\begin{equation}
D\bar{b}Db=\prod_{k}d\bar{b}_{k}db_{k}.\label{eq:Measur}
\end{equation}
In the above $\bar{b}_{i}$, $b_{i}$ are the Bose fields at lattice
site $i$ and $\bar{b}_{k}$, $b_{k}$ are their momentum-space counterparts.
Assuming condensation in the state with momentum $+k_{0}$, $b_{k_{0}}$
and $\bar{b}_{k_{0}}$ acquire non-zero mean-field values which can
be accounted for by a shift of variables,
\[
b_{i}\rightarrow\sqrt{\frac{N_{0}}{N}}e^{ik_{0}r_{i}}+b_{i},
\]
\begin{equation}
\bar{b}_{i}\rightarrow\sqrt{\frac{N_{0}}{N}}e^{-ik_{0}r_{i}}+\bar{b}_{i},\label{eq:ShiftOfBosonicFieldVriables}
\end{equation}
where $r_{i}$ stands for $i$-th lattice site position. We also assume
that the number of atoms above the condensate, $N_{B}-N_{0},$ is
small, where 
\begin{equation}
N_{0}=\bar{b}_{k_{0}}b_{k_{0}}=N_{B}-\sum_{k\neq k_{0}}\bar{b}_{k}b_{k}.\label{eq:ParticleNumberConservation}
\end{equation}
Then substituting the mean-field value Eq. (\ref{eq:ShiftOfBosonicFieldVriables})
into the action Eq. (\ref{eq:ActualBosonicAction}) and expanding
in creation/annihilation operators up to the second order the following
effective quadratic action is obtained:
\[
S_{Bog}=\intop_{0}^{\beta}d\tau\,\sum_{k}\bar{b}_{k}(\tau)\left(\partial_{\tau}+\epsilon_{B}(k)-\mu_{B}+2gn_{0}\right)b_{k}(\tau)
\]
\begin{equation}
+\frac{gn_{0}}{2}\intop_{0}^{\beta}d\tau\,\sum_{k}\bar{b}_{k_{0}+k}(\tau)\bar{b}_{k_{0}-k}(\tau)+\frac{gn_{0}}{2}\intop_{0}^{\beta}d\tau\,\sum_{k}b_{k_{0}+k}(\tau)b_{k_{0}-k}(\tau)\label{eq:BogoliubovBosonicAction}
\end{equation}
where in all the sums $k$ runs over the first BZ momenta and in the
second and third sums it is understood that if $k_{0}+k$ becomes
greater than $\pi$ its value ``wraps around'' the 1BZ edge and
becomes $k_{0}+k-2\pi$. $n_{0}$ is the condensate density, $n_{0}=N_{0}/N$,
and can be determined from the equation of state, namely the equation
that demands that the linear terms in the action vanish, $gn_{0}=\mu_{B}-\epsilon_{B}(k_{0})$.

The final step is integrating out the bosons which is made easier
by introducing the particle-hole notation:{\scriptsize{}
\begin{equation}
S_{Bog}=\intop_{0}^{\beta}d\tau\,\frac{1}{2}\sum_{k}\left(\begin{array}{cc}
\bar{b}_{k_{0}+k}(\tau) & b_{k_{0}-k}(\tau)\end{array}\right)\left(\begin{array}{cc}
\partial_{\tau}+\epsilon_{B}(k_{0}+k)-\mu_{B}+2gn_{0} & gn_{0}\\
gn_{0} & -\partial_{\tau}+\epsilon_{B}(k_{0}-k)-\mu_{B}+2gn_{0}
\end{array}\right)\left(\begin{array}{c}
b_{k_{0}+k}(\tau)\\
\bar{b}_{k_{0}-k}(\tau)
\end{array}\right),\label{eq:MatrixAction}
\end{equation}
}or, in momentum-frequency space,
\begin{equation}
S_{Bog}=\frac{1}{2}\sum_{k,\omega_{n}}\left(\begin{array}{cc}
\bar{b}_{k_{0}+k}(\omega_{n}) & b_{k_{0}-k}(-\omega_{n})\end{array}\right)G^{-1}\left(k,\omega_{n}\right)\left(\begin{array}{c}
b_{k_{0}+k}(\omega_{n})\\
\bar{b}_{k_{0}-k}(-\omega_{n})
\end{array}\right)\label{eq:MatrixActionMomentumFrequency}
\end{equation}
where
\begin{equation}
G^{-1}(k,\omega_{n})=\left(\begin{array}{cc}
i\omega_{n}+\epsilon_{B}(k_{0}+k)-\mu_{B}+2gn_{0} & gn_{0}\\
gn_{0} & -i\omega_{n}+\epsilon_{B}(k_{0}-k)-\mu_{B}+2gn_{0}
\end{array}\right).\label{eq:InverseGreensFunction}
\end{equation}
We then expand the boson-fermion interaction term Eq. (\ref{eq:BoseFermiInteractionHamiltonian})
to the first order in $\bar{b}_{k}$, $b_{k}$:
\begin{equation}
S_{BF}=U_{BF}\intop_{0}^{\beta}d\tau\,\sum_{i}\rho_{i}n_{0}+\frac{U_{BF}\sqrt{n_{0}}}{\sqrt{N}}\intop_{0}^{\beta}d\tau\,\sum_{i,k\neq k_{0}}\rho_{i}\left(\bar{b}_{k}e^{i(k_{0}-k)r_{i}}+b_{k}e^{-i(k_{0}-k)r_{i}}\right).\label{eq:FermiBoseInteractionExpanded}
\end{equation}
The first term renormalizes the fermion chemical potential, 
\begin{equation}
\mu_{F}\rightarrow\mu_{F}+U_{BF}n_{0},\label{eq:ModifiedFermionChemicalPotential}
\end{equation}
whereas the second-order terms in $\bar{b}_{k}$, $b_{k}$ (not shown
in Eq. (\ref{eq:FermiBoseInteractionExpanded})) will produce higher-order
contributions to inter-fermion interactions (three-body interactions
and higher). It is the first-order term in Eq. (\ref{eq:FermiBoseInteractionExpanded})
that produces the attractive density-density interaction between fermions.
To see this make the change of variables in the sum, $k_{0}-k\rightarrow-k$.
Then the second term of Eq. (\ref{eq:FermiBoseInteractionExpanded})
can be represented as 
\begin{equation}
\frac{1}{2}\intop_{0}^{\beta}d\tau\,\sum_{k}J^{\dagger}(k,\tau)\left(\begin{array}{c}
b_{k_{0}+k}(\tau)\\
\bar{b}_{k_{0}-k}(\tau)
\end{array}\right)+\left(\begin{array}{cc}
\bar{b}_{k_{0}+k}(\tau) & b_{k_{0}-k}(\tau)\end{array}\right)J(k,\tau)\label{eq:SourceTerm}
\end{equation}
with
\begin{equation}
J(k,\tau)=\left(\begin{array}{c}
\rho_{k}(\tau)\\
\rho_{k}(\tau)
\end{array}\right),\quad J^{\dagger}(k,\tau)=\left(\begin{array}{cc}
\rho_{-k}(\tau) & \rho_{-k}(\tau)\end{array}\right).\label{eq:SourceValues}
\end{equation}
Comparing with Eq. (\ref{eq:MatrixActionMomentumFrequency}) and employing
the rules of Gaussian integration Eq. (\ref{eq:GaussianIntegration})
we obtain the induced fermion density-density interaction
\begin{equation}
V_{Ind}(k,\omega_{n})=-\frac{U_{BF}^{2}n_{0}}{2}\frac{\epsilon_{B}(k_{0}+k)+\epsilon_{B}(k_{0}-k)-2\epsilon_{B}(k_{0})}{(i\omega_{n}+\epsilon_{B}(k_{0}+k)-\epsilon_{B}(k_{0})+gn_{0})(-i\omega_{n}+\epsilon_{B}(k_{0}-k)-\epsilon_{B}(k_{0})+gn_{0})-g^{2}n_{0}^{2}}\label{eq:InducedAttraction}
\end{equation}

\bibliographystyle{unsrtnat}

\end{document}